\documentclass[showkeys,showpacs,amssymb,preprintnumbers]{revtex4}
\usepackage{graphicx}
\usepackage{latexsym}
\usepackage{amsfonts}
\usepackage{hyperref}
\usepackage{bm}

\begin{document}

\newcommand{\vp}{\varphi}
\newcommand{\nn}{\nonumber\\}
\newcommand{\beq}{\begin{equation}}
\newcommand{\eeq}{\end{equation}}
\newcommand{\bed}{\begin{displaymath}}
\newcommand{\eed}{\end{displaymath}}
\def\bea{\begin{eqnarray}}
\def\eea{\end{eqnarray}}

%%%%%%%%%%%%%%%%%%%%%%%%%%%%%%%%%%%%%%%%%%%%%%%%%%%%%%

\title{Partition function for a singular background}
\author{J.~J.~McKenzie-Smith}
\email[Email: ]{julian.mckenzie-smith@frmhedge.com}
\affiliation{Financial Risk Management Ltd, 15 Adam Street, London WC2N 6AH, UK}
\author{W.~Naylor}
\email[Email: ]{naylor@yukawa.kyoto-u.ac.jp}
\affiliation{Yukawa Institute for Theoretical Physics, Kyoto University, 
Kyoto 606-8502, Japan}

\begin{abstract}
We present a method for evaluating the partition function in a varying external field. 
Specifically, we look at the case of a non-interacting, charged, massive scalar field at finite temperature with an associated chemical potential in 
the background of a delta-function potential. 
Whilst we present a general method, valid at all temperatures, we only give the result for the leading order term in the high temperature limit. 
Although the derivative expansion breaks down for inhomogeneous backgrounds we are able to obtain the high temperature expansion, as well as an analytic expression for the zero point energy, by way of a different approximation scheme, which we call the {\it local Born approximation} (LBA).

\end{abstract}

\pacs{11.10.Wx,  05.30.Jp}
\keywords{Quantum fields, Finite Temperature Field Theory}
\preprint{YITP-04-71}
\maketitle
\begin{center}
\today
\end{center}

%%%%%%%%%%%%%%%%%%%%%%%%%%%%%%%%%%%%%%%%

In this Letter we discuss the evaluation of the partition function in the non-smooth background of a spherically symmetric shell, a particular motivation being some recent interest in singular potentials \cite{Milton,Graet,BV}. The delta-function profile is a useful approximation in modelling semi-transparent boundary conditions, which are naturally expected in many problems of physical interest. The Casimir energy has been evaluated for a scalar field in the background of a spherical penetrable shell in \cite{Scan}, which used the Jost function technique along with contour integral methods, e.g., see \cite{Kirsten} and the references therein.

However, here we shall investigate the partition function for a massive, charged scalar field at finite temperature with a non-zero chemical potential. We shall rely on the fact that the thermal partition function can be related (by partial wave analysis) to a radial momentum integral and angular momentum sum over the phase shifts. The phase shift method has a simple connection with the Jost function approach \cite{Kirsten}. Numerous works have employed the partial waves technique to evaluate one loop and non-perturbative effects. Some of these are for example; the prefactor in bubble nucleation \cite{Lee,MN}, quantum effects for solitons \cite{FGHJ,Moss}, the Casimir energy of a skyrmion \cite{Mous} and instanton effects in QCD \cite{Hoof,Dunn}. Indeed, the phase shift method appears to have been first used in the context of field theory by Schwinger \cite{Schwing}.

Usually, the derivative expansion can be employed as an approximation to external fields that gradually vary, but for the case of potentials such as a delta-function this approximation clearly breaks down. However, recent work which deals with the 
heat kernel asymptotics of singular potentials \cite{disBV,disMoss} means that other 
approximation schemes can be devised. We shall explicitly show how by obtaining the high temperature limit using the {\it local Born approximation} (LBA).

In tandem we shall also apply the LBA to the calculation of the zero point energy, which gives surprisingly good agreement with exact results. The zero point energy must undergo regularisation; and of the many possible approaches, such as subtracting terms from the Born series \cite{FGHJ}, we tackle the problem by subtracting the relevant heat kernel coefficients, enabling an analytically continued expression for the zero point energy. As discussed in \cite{Moss} the intrinsically local heat kernel coefficients can be related to the non-local Born series. Indeed 
a non-local generalisation of the heat kernel asymptotics exists in the guise 
of covariant perturbation theory, which has been applied to finite temperature 
field theory in \cite{GZ}.

%%%%%%%%%%%%%%%%%%%%%%%%%%%%%%%%%%%%%%%%%%

As is well known, e.g., see \cite{KapBook}, the factored one loop effective action for a charged scalar field is $W^\beta=W^\beta(\mu) + W^\beta(-\mu)$, where
\begin{eqnarray}
W^\beta(\pm\mu)& = & \frac{1}{2}\log\det \ell^{2}\left[-\Box_E + m^{2} -
\mu^{2} \pm 2i\mu\frac{\partial}{\partial t}\right] 
%\nonumber \\
% +\frac{1}{2}\log\det \ell^{2}\left[-\Box_E + m^{2} -\mu^{2} - 
%2i\mu\frac{\partial}{\partial t}\right]\;.
\label{AH}
\end{eqnarray}
and the parameter $\ell$, with dimensions of 
length, keeps the argument of the determinant dimensionless. 
The connection between the Euclidean effective action and the canonical 
free energy is well known
\beq
F^\beta={1\over \beta }W^\beta
\eeq 
where we are ignoring any terms independent of the temperature, e.g., see \cite{GZ}.
We must deal with the eigenvalues of the operator 
\beq
\Delta_\pm \phi_\pm(x,\tau)= \lambda_{n,k} \phi_\pm(x,\tau)
\eeq
where
\beq
\Delta_\pm =\left[-\Box_{E\pm} + m^{2}\right] 
\eeq
and the Euclidean d'Alembertian operator is
\beq
-\Box_{E\pm}=-\left[\left(\partial_\tau\mp iA_0\right)^2+\nabla^2\right] 
\eeq
with $A_\mu=(A_0,{\bf 0})$ and $A_0 = -i\mu$.  It is related to the Lorentzian $\Box$ 
operator by the Wick rotation $-i\partial_t = \partial_\tau$.  
The vector $A_\mu$ is a fictitious Abelian gauge potential which represents the effect of an external charge 
density on the quantum system.  
%As such, it has no direct dynamical meaning, it is simply a device to incorporate the chemical potential, for a comprehensive derivation and discussion of this minimal substitution see \cite{HW}.

Consider the spatial Laplacian,  
\beq
\left(-\nabla^2 +m^2+V(x)\,\right)\phi(x)=E^2\, \phi(x)
\label{spat}
\eeq
for a set of eigenvalues $E^2$, where we shall assume a spatially dependent background coupling with spherical symmetry of the form
\beq
V(r)={\alpha\over R}\,\delta(r-R)
\label{deltpot}
\eeq 
$\alpha >0$, i.e., we are considering a massive, charged scalar field in the background of a repulsive spherical shell, at some radius $R$. We have chosen the the inverse radius for our dimensionful parameter in the background delta function. Other choices can of course be made, see \cite{Scan}. 
If we are interested in a massless field, $m=0$, a slightly different approach must be applied, e.g., see \cite{MN}. 

%The parameter $\alpha$ could also be negative definite (implying bound states) and for a constant background mass, $m$, this would model the spatial profile of the Higgs field in the thin wall limit of bubble nucleation. 

After separating the eigenmodes into radial functions and spherical harmonics, the phase shift can be obtained from the solutions of the 
$(d+1)$-dimensional radial wave equation
\beq
\left(-\partial_r^2-{d\over r}~\partial_r+m^2+{\alpha\over R}\,\delta(r-R)
+{l(l+d-1)\over r^2}\right)u(r)=E^2 \,u(r)
\eeq
with angular momentum $l$. The limit $\alpha\rightarrow\infty$ implies reflecting Dirichlet boundary 
conditions and the partial waves approach should reproduce the functional determinant for the 
Dirichlet ball which has been studied in detail in \cite{Kirsten} (however, also see the comments in \cite{Scan}). It is 
straightforward to show that the phase shift is (for $d=2$)
\bea
\label{deldis}
\tan\,\delta_{l}(k) %&=&{{\alpha\over R}~ j_{l}{}^2(k R) \over
%k\,n'_{l}(k R)\,j_{l}(k R) 
%- k\, j'_{l}(k R)\,n_{l}(k R)+{\alpha\over R}~ j_{l}{}(k R)n_{l}{}(k R) }
%\\
&=&{\frac \pi 2\, \alpha \,J_{l+1/2}^2(k R) \over
\frac \pi 2\, \alpha \,J_{l+1/2}{}(k R)N_{l+1/2}{}(k R)-1 }
\eea
where the radial momentum is defined by
\beq
k=\sqrt{E^2-m^2}
\eeq
and $J_{\nu}(x),~N_{\nu}(x)$ are Bessel and Neumann functions respectively.
%The above result is the one quoted for three spacial dimensions ($d=2$). 
For such a case when the eigenvalues are not explicitly known it is possible 
to employ the phase shift method as follows. 
%(Expanding the phase shift in terms of small 
%$\alpha$ gives the Born approximation to each order and hence the non-local 
%heat kernel coefficients.)

%%%%%%%%%%%%%%%%%%%%%%%%%%%%%%%%%%%%%%%%

From a knowledge of the phase shift it is straight forward to relate it 
to the heat kernel by
\bea
\label{ker}
K(t) %&=& n_0+\frac{2}{\pi} \int_{0}^{\infty}dk\:e^{-(k^2+m^2) t}\,k\,t
%\sum_l\:\chi_l\left(\delta_l(k)-\delta_l(0)\right)\nn
&=&\frac{2}{\pi} \int_{0}^{\infty}dk\:e^{-(k^2+m^2) t}\,k\,t
\sum_l\:\chi_l\delta_l(k)
\eea
e.g., see \cite{MN,Moss}. We should mention that in this Letter we shall assume that there are no bound states, which is the case for background coupling with $\alpha > 0$ (e.g., see \cite{Lee,Moss} for a discussion of the inclusion of any bound states). The degeneracy factor $\chi_l=(2l+1)$ in three dimensions. 
There is also a free space contribution, which is encoded by the $C_0$ term in the heat kernel expansion. This term gives the usual constant background results, which have been well studied at finite temperature \cite{KapBook,HW}, so we shall not discuss it.  
The heat kernel can now be used to regularise the one loop effective action.
Defining the generalised $\zeta$-function \cite{DC} by
\begin{equation}
\zeta(s)= {1\over \Gamma(s)}
\int_{0}^{\infty}\,t^{s-1}\, {\rm tr}%\,\left( K(t)-n_0\right)
~K(t)~dt
\label{zeta}
\end{equation}
The analytic continuation of $\zeta(s)$ then gives the one loop effective action, which is related to the zeta-function by
\begin{equation}
\label{forapen}
W^\beta=W^\beta(\mu) +W^\beta(-\mu)\qquad\qquad W^\beta(\pm\mu)=-\frac 1 2\zeta_\pm'(0)+\frac 1 2\zeta_\pm(0)\log\ell^2
\end{equation}

Because we are working on a static manifold, the field can be written in separable form as $\phi(\tau,x)=e^{-i\omega_n \tau} \phi(x)$, where $\omega_n=2\pi n/\beta$ are the Matsubara frequencies with $\beta$ the inverse temperature.  Thus, the coincidence limit of the heat kernel takes the following form
\begin{equation}
K^{\beta}_\pm(t)=
\sum_{n=-\infty}^{\infty}
 \frac{2}{\pi} \int_{0}^{\infty}dk\:e^{-(k^2+m^2)t}\,k\, t
\sum_l\:\chi_l\delta_l(k)
 e^{-(\omega_n\pm i \mu)^2 t}
\end{equation}
The thermal effective action for a charged scalar field is therefore
\begin{equation}
\label{frbs}
W^\beta(\pm\mu)=-\frac{1}{2}\int_{0}^{\infty}dt~t^{s-1}
\sum_{n=-\infty}^{\infty}
 \frac{2}{\pi} \int_{0}^{\infty}dk\:e^{-(k^2+m^2)t}\,k\,t
\sum_l\:\chi_l\delta_l(k)
 e^{- (\omega_n\pm i \mu)^2 t}
\end{equation}
The analytic continuation can be conveniently performed by using the Jacobi-Poisson resummation formula on the Matsubara modes, e.g., see \cite{WW}. After some formal manipulations we find the total thermal effective action to be
\bea\label{bosew}
W^\beta&=&-\frac{\beta}{\pi}\int_{0}^{\infty}~dk
\sum_{l=0}^{\infty}(2l+1)\:\bar{\delta}_{l}(k)
		 -\frac{\beta}{\pi}\int_{0}^{\infty}{k\ dk\over (k^2+m^2)^{1/2}}
\sum_{l=0}^{\infty}(2l+1)\:\frac{\delta_{l}(k)}{e^{\beta (E-\mu)}-1}
-\big(\mu\rightarrow -\mu\big)\nn
&=&W^{\infty}+W^\beta(\mu)+W^\beta(-\mu)
\eea
In an abuse of our previous notation, see (\ref{forapen}), the first term is the zero point energy 
($\beta\rightarrow\infty$) which is ultraviolet divergent, hence
$\bar{\delta_{l}}(k)$ (see below), while the other two terms contribute to the thermal part of the effective action, one for each charge $\pm\mu$. 
Similar expressions can be found in \cite{Lee,MN}, however, here we have 
incorporated a chemical potential.

%%%%%%%%%%%%%%%%%%%%%%%%%%%%%%%%%%%%%%%%%%%%%%%%%

For numerical purposes, the analytic continuation of the zero point energy 
is best performed by subtracting terms from the heat kernel. As $t\to 0$, the heat kernel in $d+1$ dimensions has the asymptotic expansion
\begin{equation}\label{kerexp}
K(t) \sim  t^{-(d+1)/2}\,\sum_{n=0}\,C_{n}(r)t^{n}
\end{equation}
where due to radial symmetry the heat kernel coefficients are only local 
functions of $r$.
The leading terms, which cause the poles in the zeta-function, can
be removed by replacing the sum over phase shifts in (\ref{ker}) by \cite{MN}
\bea
\label{reg}
\sum_l\,\chi_l\bar{\delta}_{l}(k)=
\sum_l\,\chi_l\delta_{l}(k) &-& \frac{\pi
C_{1}(r)\,(k^2+m^2)^{d-1\over 2}}{\Gamma(\frac{d+1}{2})}
- \frac{\pi C_{3/2}(r)\,(k^2+m^2)^{d-2\over 2}}{\Gamma(\frac{d}{2})}\nn
&-&\frac{\pi C_{2}(r)\,(k^2+m^2)^{d-3\over 2}}{\Gamma(\frac{d-1}{2})}
- \frac{\pi C_{5/2}(r)\,(k^2+m^2)^{d-4\over 2}}{\Gamma(\frac{d-2}{2})}
\eea
where for distributional sources the heat kernel coefficients only contribute to surface terms \cite{disBV,disMoss}, see below. 
To obtain a finite regularised sum we only require up to $C_2$ for finite temperatures in three dimensions and zero temperature in four dimensions. However, in practice it is useful to include the $C_{5/2}$ term for better numerical convergence or to interpolate the large $k$ behaviour. A useful check is that the leading asymptotic behaviour follows that of the $C_{5/2}$ term. Indeed, as we shall show, the dominant contribution to the LBA is given by the $C_{5/2}$ term. 

The analytic continuation of the zero point energy can be found in \cite{Mous,MN} 
for example, and we obtain
\bea
W^{\infty}&=&-\frac{\beta}{\pi} \int_{0}^{\infty}dk
\left(\:\sum_{l=0}^{\infty}(2l+1)\delta_{l}(k)-2\sqrt{\pi}\sqrt{k^2+m^2}~C_{1}
-\pi ~C_{3/2}
-\frac{\sqrt{\pi}~C_{2}}{\sqrt{k^2+m^2}}
-\frac{\pi ~C_{5/2}}{k^2+m^2}\right)\nn
&& +\frac{\pi\beta ~C_{5/2}}{2m}+ \frac{\beta ~C_{2}}{\sqrt{4\pi}}\log m^{2}\ell^2
\label{ntherm}
\eea
In Fig. 1 is a plot of the integrand (in brackets) in the above equation, 
given as blue solid lines for various values of the coupling $\alpha$.  
The extra factor of two for the zero point energy (\ref{ntherm}) as compared to the expression given in \cite{MN} is due to the fact that we are considering a charged scalar field. Although in this Letter we do not give explicit results that depend on $\mu$, apart from (\ref{bosew}), the more general expression is required if one wishes to comment on things such as Bose-Einstein condensation, which we hope to report on in the near future.

For the spatial wave equation (\ref{spat}) with a potential of the form 
defined in (\ref{deltpot}) the heat kernel coefficients are \cite{disBV,disMoss}:
\begin{eqnarray}
C_{1}(r)={1\over (4\pi)^{(d+1)/2} }\int_\Sigma -{\alpha\over R}
\qquad\qquad &&
C_{3/2}(r)={1\over (4\pi)^{(d+1)/2} } \int_\Sigma {\sqrt\pi\over 4} 
{\alpha^2\over R^2}
\label{c132}\\
C_{2}(r)={1\over (4\pi)^{(d+1)/2} }\int_\Sigma \left(-\frac1 6 {\alpha^3\over R^3}
+m^2{\alpha\over R}\right)\qquad\qquad &&
C_{5/2}(r)={1\over (4\pi)^{(d+1)/2} }\int_\Sigma 
\left({\sqrt\pi\over 32} {\alpha^4\over R^4}+
{\sqrt\pi\over 4} m^2 {\alpha^2\over R^2}\right)
\label{c52}
\end{eqnarray}
where the volume of a spherical shell in $d+1$ dimensions is given by 
\beq
\int_\Sigma={2\pi^{(d+1)/2}R^{d}\over \Gamma({d+1\over 2})}
={4\pi R^2}\Big|_{d=2}
\eeq

In fact, the leading order contribution comes from the $C_{5/2}$ term in (\ref{ntherm}), up to the renormalisation scale dependence, $\ell$. 
As can be seen in Fig. 1, for example, the choice $\alpha={x\over m R}$ for various values of $x$ gives surprisingly good agreement with the integrand of (\ref{ntherm}). That is, the 
the zero point energy is dominated by the term
\beq
W^\infty\approx\frac{\beta}{\pi} \int_{0}^{\infty}dk~
\left(\frac{\pi~C_{5/2}}{k^2+m^2}\right)=\frac{\pi\beta ~C_{5/2}}{2 m}
\label{domin}
\eeq
for which the integrand of the left hand side (in brackets) we have plotted in Fig. 1 as red dashed lines.
Clearly, the accuracy of this approximation improves as the area of the integrand in (\ref{ntherm}) becomes larger and the oscillatory behaviour for the large $k$ part has less effect. Including higher order corrections, such as the $C_3$ term, should improve the accuracy for smaller values of the area of the integrand in (\ref{ntherm}).

%%%%%%%%%%%%%%%%%%%%%%%%%%%%%%%%%%%%%%%%%%%%%%
\begin{figure}[tbh]
\begin{center}
\scalebox{0.7}{\includegraphics{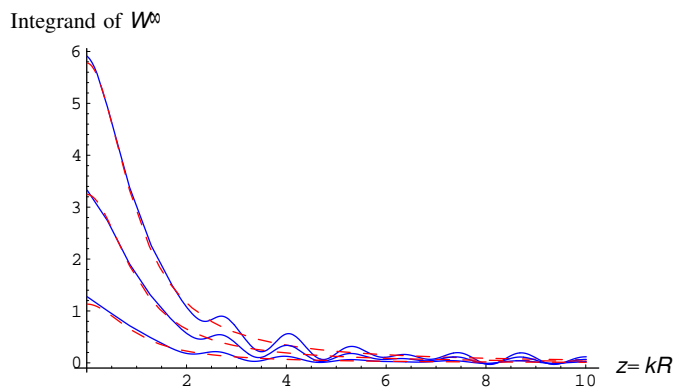}}
\end{center}
\caption{ 
A plot of the integrand in (\ref{ntherm}), blue lines, and the contribution from the integrand in (\ref{domin}), red dashed lines, for values of increasing $\alpha=1.5,~2.25,~\&~2.75$, in units of $(m R)^{-1}$. For clarity the $z$ axis has been shifted below the origin by a small amount.
\label{plot}} 
\end{figure}
%%%%%%%%%%%%%%%%%%%%%%%%%%%%%%%%%%%%%%%%%%%%%%%%%%%%%%%

For large temperatures numerical evaluation of $W^\beta(\pm\mu)$ in (\ref{bosew}) becomes difficult and  it is useful to have analytic expressions valid in the high temperature limit, $\beta\rightarrow 0$. We can achieve this by using the LBA, as we have already done for the zero point energy. 
The high temperature expansion can be obtained in various ways, however, we shall simply adapt the method used in \cite{HW}. First, let's rewrite the thermal effective action explicitly in terms of $k$ as
\begin{equation}
W^{\beta}(\pm\mu)=-\frac{\beta}{\pi}\int_{0}^{\infty}{k\ dk\over (k^2+m^2)^{1/2}}
\sum_{l=0}^{\infty}(2l+1)\:\frac{\delta_{l}(k)}
{e^{\beta(k^2+ m^2)^{1/2}}~e^{\pm\beta\mu}-1}
\label{therm}
\end{equation}
Our approach will be to substitute the sum over phase shifts \cite{MN}
\beq
\sum_l\,\chi_l\delta_{l}(k) \sim  \sum_{n=1}^\infty {\pi~
C_{n}\,(k^2+m^2)^{{d+1-2n\over 2}}\over \Gamma(\frac{d+1-2n+2}{2})}
\label{sumov}
\eeq
into our expression for $W^\beta$ (\ref{therm}), i.e., 
$W^\beta(\pm r)=\sum_n W^{\beta}_n(\pm r)$
\beq
W^{\beta}_n(\pm r)=
%\sum_{n=1}^\infty 
{-\beta^{2n-d-1}~C_{n}\,\over \Gamma(\frac{d+1-2n+2}{2})}
\int_{0}^{\infty} x~dx \frac{(x^2+\bar m^2)^{{d-2n}\over 2}}
{e^{(x^2+\bar m^2)^{1/2}}~e^{\pm r\bar m}-1}
\eeq
where we have made the change of variables $x=\beta k$,~$\bar m=\beta m $ and $r=\mu/ m$.

In this Letter we shall only consider the leading order contribution to 
the high temperature expansion (for $d=2$), which corresponds to $n=1$, i.e., 
\beq
W^{\beta}_1(\pm r)=-\beta^{-1}
 {C_{1}\,\over \Gamma(\frac{3}{2})}
\int_{0}^{\infty} x~dx\frac{1}{e^{(x^2+\bar m^2)^{1/2}}~e^{\pm r \bar m}-1}
\eeq
The small $\bar m$ expansion of the above integral is well known \cite{HW} and leads to the result
\beq
W^{\beta}_1(\pm r)={-2 C_1\over \sqrt\pi\beta}~g_2(\bar m,\pm r)
\eeq
where the function $g_2(\bar m,r)$ is defined in Appendix A of \cite{HW}. 
Then, it is simple to show that 
\beq
g_2(\bar m,r)+g_2(\bar m,-r)={\pi^2\over 3}-2\bar m+ \bar m r\ln \left[{(1+r)\over (1-r)}\right]+{\bar m^2 (1-r^2)\over 2}
+{\cal O}(\bar m^3)
%\sum_{k=1}^\infty {(-1)^k \over k}~\zeta(2k)~
%\left(\bar m(r-1)+\bar m(-r-1)\over 2\pi\right)^{2k}
\eeq
and thus, the leading order contribution for small $\bar m$ to the thermal 
part of the effective action is 
\beq
W^{\beta}_1(\mu)+W^{\beta}_1(-\mu)=\left({-2 C_1\over \sqrt\pi\beta}\right)
{\pi^2\over 3}=
{1\over 12\beta}\int_\Sigma {\alpha\over R}=
{1\over 12\beta}~{4\pi\alpha R}
\eeq
where the second equality is specific to a spherical shell. Thus, for the free space field theory, i.e., in a constant background, the leading order contribution to the free energy is $F_0^\beta\propto \beta^{-4}$, e.g., see \cite{HW}, whereas the effect of a repulsive delta-function potential is to contribute to leading order $F_1^\beta\propto \beta^{-2}$.

%%%%%%%%%%%%%%%%%%%%%%%%%%%%%%%%%%%%%%%%%

The method we have discussed is sufficiently general such that we can choose any spherically symmetric background potential, smooth or inhomogeneous, the only requirement being a knowledge of the heat kernel coefficients. Of course if they are not known then the heat kernel coefficients can be derived by taking the local part of the Born approximation, which is an expansion in powers of small coupling of the phase shift. This is also true for the case when there is no analytic expression for the phase shift, where recourse can be made to the WKB method \cite{Lee,Dunn}.

Furthermore, the finite temperature expression (\ref{therm}) is a general expression valid at any temperature, which requires numerics for the general case. However, it is instructive to have analytic expressions for a charged scalar field in the high temperature limit, for example to consider Bose-Einstein condensation. In this Letter 
we showed how to obtain the leading order term. Indeed, next to leading order corrections will depend on the chemical potential, $\mu$.  
The low temperature expansion can also be obtained in a similar way.

The {\it local Born approximation} (LBA) was also applied to give an approximate expression for the non-thermal part of the one loop effective action, i.e., the 
zero point energy. 
We employed zeta-function and heat kernel methods to subtract the divergences inherent in the zero point energy, this being one of the many commonly used subtraction procedures.

As well as for delta-function type potentials one could consider step function profiles, which are useful in modelling many physically interesting situations, such as instantons in bubble nucleation \cite{MN}. For such a case it would also be possible for the chemical potential, $\mu$, to vary with the same radial profile as the mass, e.g., a step-function with profiles $\mu^2\theta(r-R)$ and $m^2\theta(r-R)$. However, this is a considerably more complicated set up, which we shall report on in forthcoming work, as well as other related issues.

%%%%%%%%%%%%%%%%%%%%%%%%%%%%%%%%%%%%%%%
\section*{Acknowledgments}
We would like to thank S.~G.~Cox for fruitful discussions during the 
initial stages of this work and also the referee for useful comments.
W.~N. is supported by the 21COE program ``Centre for Diversity and Universality in Physics,'' {\bf @} Kyoto University.

%%%%%%%%%%%%%%%%%%%%%%%%%%%%%%%%%%%%%%%%%%
%\newpage

%%%%%%%%%%%%%%%%%%%%%%%%%%%%%%%%%%%%%%%
\end{document}